\documentclass[prd,twocolumn,reprint,preprintnumbers,nofootinbib,superscriptaddress]{revtex4-2}
\input{header.tex}

\reportnum{-2}{\qquad\qquad CERN-TH-2025-215}
\reportnum{-3}{\qquad\qquad P3H-25-083}
\reportnum{-4}{\qquad\qquad TTP25-039}

\newcommand{\nc}{N_{c_D}} 
\newcommand{\nf}{N_{f_D}}     
\newcommand{\pid}{{\pi_D}}   
\newcommand{\rhod}{{\rho_D}}   
\newcommand{\mpi}{{m_{\pid}}}     
\newcommand{\mrho}{{m_{\rhod}}}   
\newcommand{\fpi}{f_{\pi_D}}    
\newcommand{\ed}{e_D}      
\newcommand{\Ld}{\Lambda_D}     
\newcommand{\qd}{q_D}           

\usepackage{colortbl}
\definecolor{LightBlue}{rgb}{0.91,0.96,1}
\usepackage[dvipsnames]{xcolor}

\begin{document}

\title{Sub-GeV dark matter and multi-decay signatures \\from dark showers at beam-dump experiments}

\author{Elias Bernreuther}
\email{ebernreuther@ucsd.edu}
\affiliation{Department of Physics, University of California, San Diego, La Jolla, CA 92093, USA}

\author{Nicoline Hemme}
\email{ nicoline.hemme@kit.edu}
\affiliation{Institute for Theoretical Particle Physics (TTP), Karlsruhe Institute of Technology (KIT), 76128 Karlsruhe, Germany}

\author{Felix Kahlhoefer}
\email{kahlhoefer@kit.edu}
\affiliation{Institute for Astroparticle Physics (IAP), Karlsruhe Institute of Technology (KIT), Hermann-von-Helmholtz-Platz 1, 76344 Eggenstein-Leopoldshafen, Germany}

\author{Suchita Kulkarni}
\email{
suchita.kulkarni@uni-graz.at}
\affiliation{Institute of Physics, NAWI Graz, University of Graz, Universit\"atsplatz 5, A-8010 Graz, Austria}

\author{Maksym Ovchynnikov}
\email{maksym.ovchynnikov@cern.ch}
\affiliation{Theoretical Physics Department, CERN, 1211 Geneva 23, Switzerland}

\begin{abstract}
In models of strongly interacting dark sectors, the production of dark quarks at accelerators can give rise to dark showers with multiple dark mesons in the final state. If some of these dark mesons are sufficiently light and long-lived, they can be detected with searches for displaced vertices at beam-dump experiments and electron-positron colliders. In this work we focus on the case that dark quark production proceeds via effective operators, while the dark sector analogue of the $\rho^0$ meson can decay via kinetic mixing. We evaluate current constraints from NA62 and BaBar as well as sensitivity projections for SHiP and Belle II. We find that there exists a sizeable parameter region where SHiP may detect several displaced vertices in a single event and thus obtain valuable information about the structure of the dark sector.
\end{abstract}

\maketitle

\section{Introduction}

Dark sector models are based on the idea that there may exist particles that interact strongly with each other but only very weakly with visible matter. A natural way to realise such a set-up is to consider a non-Abelian gauge extension of the Standard Model (SM) with asymptotic freedom~\cite{Kribs:2016cew}. At low energies, these theories feature chiral symmetry breaking,  and the fermions participating in these interactions, called dark quarks, confine into bound states called dark mesons. Since SM particles are not charged under the new gauge group, a separate mechanism is needed to mediate interactions between the two sectors, such as renormalisable portal interactions or non-renormalisable effective operators.

Rapidly growing interest in such strongly interacting dark sectors has been triggered by two key observations. The first is that some of the dark mesons can be stable and provide viable dark matter candidates~\cite{Bai:2010qg,Buckley:2012ky,Cline:2013zca}. Conversion and number-changing processes within the dark sector then may be responsible for setting the dark matter relic abundance~\cite{Hochberg:2014dra,Hochberg:2014kqa,Choi:2018iit,Beauchesne:2018myj,Bernreuther:2019pfb,Bernreuther:2023kcg,Chu:2024rrv,Appelquist:2024koa,Garcia-Cely:2024ivo,Garcia-Cely:2025flv}, providing an alternative mechanism that evades many of the constraints putting pressure on Weakly Interacting Massive Particles. The second observation is that the unstable dark mesons predict exciting signatures at collider experiments~\cite{Strassler:2006im,Daci:2015hca,Pierce:2017taw,Kribs:2018ilo,Cheng:2019yai,Butterworth:2021jto,Knapen:2021eip,Cheng:2024hvq}. If a pair of highly-energetic dark quarks is produced in a collision of visible particles, we expect hadronisation and fragmentation processes leading to a high multiplicity of final states. If the dark sector behaves in a way similar to QCD, the result will be a dark shower with a mixture of stable and decaying dark mesons, leading to so-called semi-visible jets~\cite{Cohen:2015toa,Beauchesne:2017yhh} (or emerging jets~\cite{Schwaller:2015gea} if the decaying dark mesons are long-lived).

Most studies in the literature focus on only one of these two aspects of strongly interacting dark sectors.\footnote{See, however, Ref.~\cite{Berlin:2018tvf} for a notable early exception.} The reason is that models that address the dark matter puzzle are typically difficult to detect at the LHC because they predict a large fraction of stable (and hence invisible) dark mesons and small masses for the unstable dark hadrons (at or below the GeV scale). However, the LHC is not the only experiment capable of producing dark showers. In a recent study~\cite{Bernreuther:2022jlj}, some of us pointed out the possibility of searching for dark showers also at Belle II. Since then, the approval of the SHiP experiment at CERN~\cite{Alekhin:2015byh,SHiP:2025ows} has opened up an exciting new avenue for exploring dark sectors at the GeV scale, offering unprecedented discovery potential for many types of light and long-lived particles.

The main goal of the present work is to determine the sensitivity of SHiP for GeV-scale strongly interacting dark sectors. Specifically, we focus on a set-up with two types of dark mesons. The dark pions, arising as pseudo-Nambu-Goldstone bosons of chiral symmetry breaking, are stable and constitute the dark matter candidate. The dark rho mesons, on the other hand, are predicted to be heavier and decay via kinetic mixing with SM photons~\cite{Holdom:1985ag}. Motivated by astrophysical and cosmological arguments~\cite{Bernreuther:2023kcg}, we consider the mass hierarchy $m_{\rhod} < 2 m_{\pid}$, for which the dark rho mesons decay exclusively into visible final states. For sufficiently small couplings, their decay length may be macroscopic, giving rise to displaced vertices~\cite{Albouy:2022cin}, which can be efficiently targeted at SHiP.

The phenomenology of dark rho mesons closely resembles that of dark photons, the massive gauge bosons of a broken Abelian gauge symmetry~\cite{Pospelov:2007mp,Ilten:2018crw,Fabbrichesi:2020wbt,Kyselov:2024dmi}. Consequently, in case of detection, the standard signature -- a single decay vertex -- has little discriminating power between the two models. The key difference is that in minimal dark photon models (where interactions arise solely from mixing with the SM photon), the probability to produce two dark photons in a single event is negligible, whereas in models of strongly interacting dark sectors, the probability to produce multiple dark rho mesons in a single event can be quite large.  Observing several displaced vertices in a single collision would therefore immediately exclude the minimal dark-photon scenario. Moreover, differentiation from non-minimal dark-photon models with multiple decays per event (see e.g.\ Refs.~\cite{Schabinger:2005ei,Curtin:2013fra,Cheng:2024gfs}) can be achieved by considering additional kinematic variables, such as the total invariant mass of the decaying system~\cite{DallaValleGarcia:2025aeq}.

In this work, we focus on the case where the dark quarks couple to visible particles via an effective dimension-6 operator, which arises from integrating out a heavy dark photon with kinetic mixing. Once the dark sector gauge group and number of dark quark flavours are fixed, this set-up yields a very economical parameter space comprising only the dark pion mass, the dark rho meson mass, and the suppression scale of the effective interaction, while all other properties of the dark sector are inferred from lattice simulations. We note here that the number of dark colours ($\nc$) and flavours ($\nf$) are restricted to $\nf/\nc \ll 3$ to be in the chirally-broken phase~\cite{Albouy:2022cin,Kulkarni:2025rsl} and the dark meson mass spectrum may depend on this ratio~\cite{Alfano:2025non}. We evaluate constraints from past and ongoing beam-dump experiments, including both perturbative and non-perturbative production modes for dark rho mesons. Moreover, we update the constraints from displaced-vertex searches at BaBar~\cite{BaBar:2015jvu} and the sensitivity projections for Belle II previously derived in Ref.~\cite{Bernreuther:2023kcg} using improved dark shower simulations. We find that SHiP can probe large regions of unexplored parameter space. In a sizable part of these parameter regions, SHiP is predicted to observe events with multiple displaced vertices.

The remainder of this work is structured as follows. In section~\ref{sec:set-up} we describe the dark sector that we study and its interactions with the SM. Section~\ref{sec:phenomenology} then focuses on the phenomenological implications, in particular the different production modes for dark rho mesons. We describe the experiments under consideration and our analysis in section~\ref{sec:analysis} before presenting our results in section~\ref{sec:results}. Our conclusions are presented in section~\ref{sec:conclusion}. Additional details on the signal simulation are provided in appendices~\ref{app:DetailsOnTheSimulatedSample} and~\ref{app:event-calc}.

\section{Model set-up}
\label{sec:set-up}

We consider an $SU(3)\times U(1)'$ gauge extension of the SM with two Dirac fermions $\qd$, referred to as dark quarks, each transforming in the fundamental representation of $SU(3)$ and carrying opposite charge $\pm 1$ under $U(1)'$. We assume mass-degenerate dark quarks with a small mass, $m_{\qd}\lesssim\Ld$, where $\Ld$ is the dark confinement scale at which the $SU(3)$ gauge coupling becomes strong and the theory confines. 

At energies below $\Ld$, the approximate global chiral symmetry $SU(2)_L\times SU(2)_R\times U(1)'$ is spontaneously broken to $SU(2)_V \times U(1)'$. This symmetry breaking gives rise to three pseudo-Nambu-Goldstone bosons known as dark pions, $\pi_D^+$, $\pi_D^0$, and $\pi_D^-$, with $U(1)'$ charge $+2$, 0, and $-2$, respectively. These pseudoscalar mesons are the lightest bound states in the spectrum, while the next-lightest bound states are the vector mesons, $\rho_D^+$, $\rho_D^0,$ and $\rho_D^-$. In the following, we will be particularly interested in the neutral dark rho meson $\rho_D^0$, which we will denote as $\rhod$ and refer to as the dark rho meson for simplicity. 

Due to the conservation of $U(1)'$ charge, the charged dark pions are predicted to be stable. In a theory with two dark flavours, also the neutral dark pions can be stabilised by an appropriate discrete symmetry, making the dark pions viable candidates for dark matter~\cite{Bai:2010qg,Buckley:2012ky,Cline:2013zca,Bernreuther:2023kcg}. The relic abundance of dark pions can, in principle, be set by the thermal freeze-out of the number-changing process $3\pid\rightarrow2\pid$. Bounds on the dark matter self-interaction cross section from the Bullet Cluster, however, imply that this process is insufficient to deplete the dark matter abundance. This problem is solved in the presence of dark rho mesons with $\mrho < 2 \mpi$, which can deplete the dark pion abundance via the conversion process $3\pid\rightarrow\pid\rhod$~\cite{Bernreuther:2023kcg}, provided that the dark rho mesons can decay into SM particles sufficiently fast.

Such interactions are provided by the $U(1)'$ gauge group, which introduces a new massive gauge boson $Z'$ that can kinetically mix with the SM hypercharge even if none of the SM fermions carry $U(1)'$ charge. We will focus on the case that the $Z'$ is heavy compared to the energy scale of the process, which for SHiP implies $m_{Z'} \gg \sqrt{s} \approx 27.4 \, \mathrm{GeV}$. 
The interactions between dark quarks and SM fermions are then given by the effective operator~\cite{Bernreuther:2022jlj}
\begin{equation}
    \mathcal{L}_{\rm eff} = \frac{1}{\Lambda_\text{eff}^{2}}\bar{q}_{D}\gamma^{\mu}\qd J_{\mu,\text{EM}},
    \label{eq:interactions}
\end{equation}
where $J_{\mu,\text{EM}}$ is the electromagnetic current and $\Lambda_\text{eff} = m_{Z'}/\sqrt{\kappa e \ed}$ with the kinetic mixing parameter $\kappa$, and $e$ and $\ed$ denoting the gauge coupling of $U(1)_\text{EM}$ and $U(1)'$, respectively.

Below the dark confinement scale, we can use the one-particle replacement~\cite{Bernreuther:2019pfb,Klingl:1996by}
\begin{equation}
\bar{q}_{D}\gamma^{\mu}q_{D} \to \frac{2m_{\rhod}^2}{g_{\pi \rho}}\rho_D^{\mu},
\label{eq:1-particle-replacement}
\end{equation}
where $g_{\pi\rho}$ denotes the coupling between dark pions and dark rho mesons. The KSRF relation~\cite{Kawarabayashi:1966kd,Riazuddin:1966sw} implies $g_{\pi\rho} \approx m_\rhod/(\sqrt{2}\fpi)$ with the dark pion decay constant $\fpi$. The interaction between a single dark rho meson and SM particles can therefore be written as
\begin{equation}
    \mathcal{L}_{\rm eff}= \frac{2 m_{\rhod}^{2}}{\Lambda_\text{eff}^{2} \, g_{\pi\rho}}
    \cdot {\rho_D^\mu}J_{\mu,\text{EM}} \; .
    \label{eq:interaction-one-rho}
\end{equation}
This interaction looks exactly\footnote{Note that in our set-up we always vary $\mrho$ and $\Ld$ simultaneously and therefore never encounter the case that $\mrho \gg \Ld$, in which the vertex of a composite $\rhod$ would be suppressed based on constituents counting basis~\cite{Lepage:1980fj}.} like that of a dark photon, $\mathcal{L} \supset e \epsilon \rho_D^{\mu} J_{\mu,\text{EM}}$,
with the effective mixing angle
\begin{equation}
    \epsilon = \frac{2 m_{\rhod}^{2}}{\Lambda_\text{eff}^{2} \, g_{\pi\rho}e} \; .
    \label{eq:epsilon}
\end{equation} In other words, the dark rho meson inherits the production and decay modes of a dark photon~\cite{Ilten:2018crw,Kyselov:2024dmi,Kyselov:2025uez}.
For $\mrho<2\mpi$, the $\rhod$ is kinematically forbidden from decaying into dark pions and therefore decays exclusively to SM states. This can lead to macroscopic decay lengths of the dark rho meson if $\mrho / \Lambda_\text{eff}$ is sufficiently small.

At the same time, $\mrho$ cannot be much smaller than $2 \mpi$ for the theory to remain within the validity range of chiral perturbation theory. We will therefore focus on two specific mass ratios $r \equiv \mrho / \mpi$, namely $r = 1.5$ and $r = 1.9$. For these mass ratios, we can determine $\Ld / \mpi$ and $\fpi / \mpi$ from lattice simulations~\cite{Maris:2005tt}, see Ref.~\cite{Albouy:2022cin}. These results also enable us to infer the dark pion-rho coupling, which is found to be approximately $g_{\pi\rho} \approx 5.7$. Given these inputs and a specific value of $r$, our model is fully described by only two independent parameters: the dark rho meson mass $\mrho$ and the suppression scale $\Lambda_\text{eff}$. For reasons that will be discussed in more detail below, we will focus on the parameter ranges $200 \, \mathrm{MeV} \leq \mrho \leq 5 \, \mathrm{GeV}$ and $100 \, \mathrm{GeV} \leq \Lambda_\text{eff} \leq 100 \, \mathrm{TeV}$.

\section{Phenomenology of dark rho mesons}
\label{sec:phenomenology}

To calculate the lifetime and branching ratios of dark rho mesons, we can directly apply the well-known results for dark photons, making use of eq.~\eqref{eq:epsilon} to translate between the different parameters. In particular, the partial decay width into hadrons is inferred from the hadronic cross section ratio $R$, which features resonant peaks around the $\omega$ and $\phi$ meson masses~\cite{ParticleDataGroup:2024cfk}.

For the production of dark rho mesons, on the other hand, we need to consider both the conventional production modes of dark photons and the production mode specific to our model, which is the hadronisation of dark quarks into dark rho mesons in dark showers. In terms of experimental signatures, the latter mode is qualitatively different from the dark-photon-like production modes, because it may produce more than one dark rho meson per event. Hence, it allows for the differentiation between dark photons and dark rho mesons, as we will discuss in detail in Sec.~\ref{sec:analysis}. Let us begin with the discussion of dark showers and then turn to the other production modes.

\subsection{Dark shower production}

\begin{figure*}
    \centering
    \includegraphics[width=0.32\textwidth]{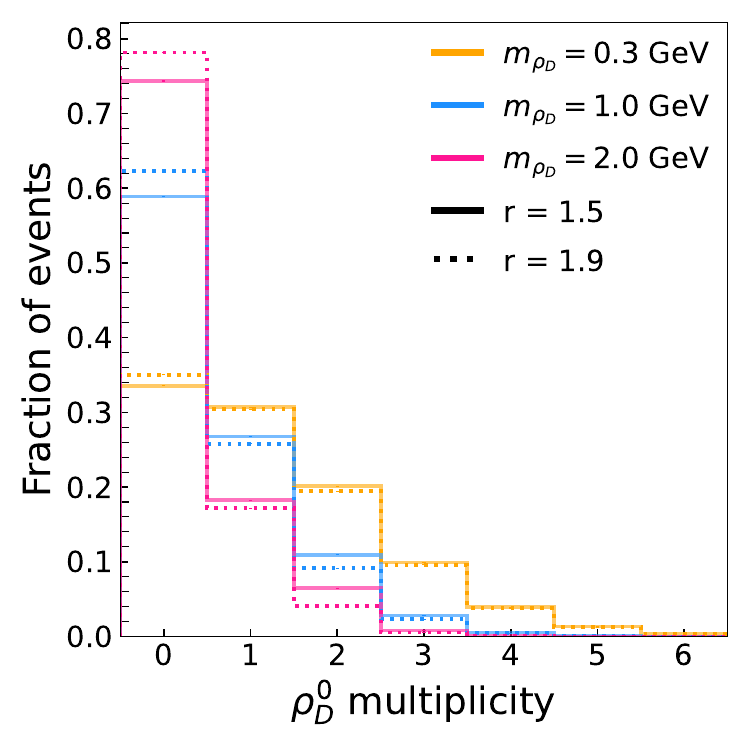}
    \includegraphics[width=0.32\textwidth]{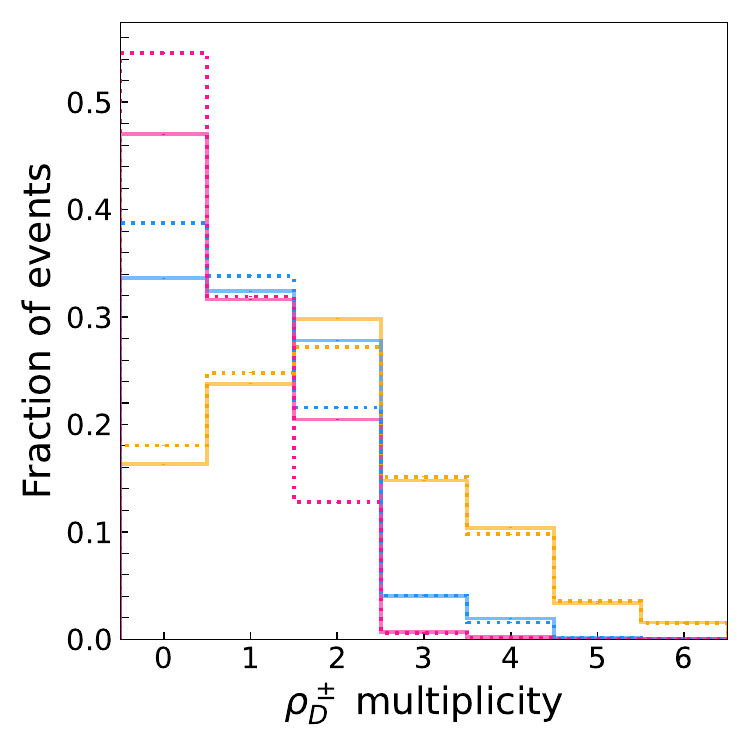}
    \includegraphics[width=0.32\textwidth]{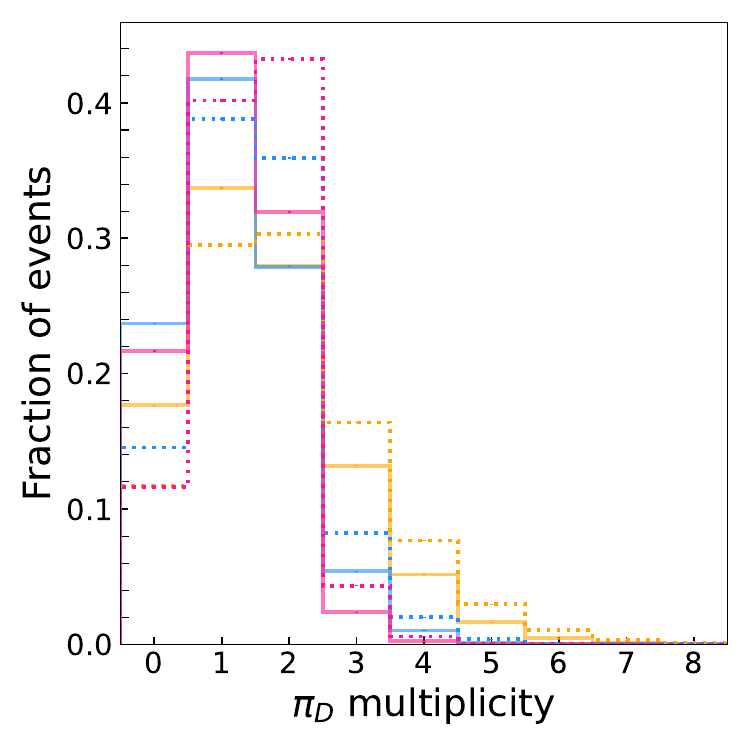}
  \vspace{-3mm}
    \caption{Multiplicity distribution of the various dark mesons predicted for the SPS beam-dump facilities. The panel shows the multiplicity distribution of neutral dark rhos (left), charged dark rhos (middle), and charged and neutral dark pions (right). The distributions are normalised to unity. Error bars represent standard Poisson uncertainties of the Monte Carlo samples.}
    \label{fig:multiplicities}
\end{figure*}

\begin{figure*}[t]
 \centering
  \includegraphics[width=0.5\textwidth]{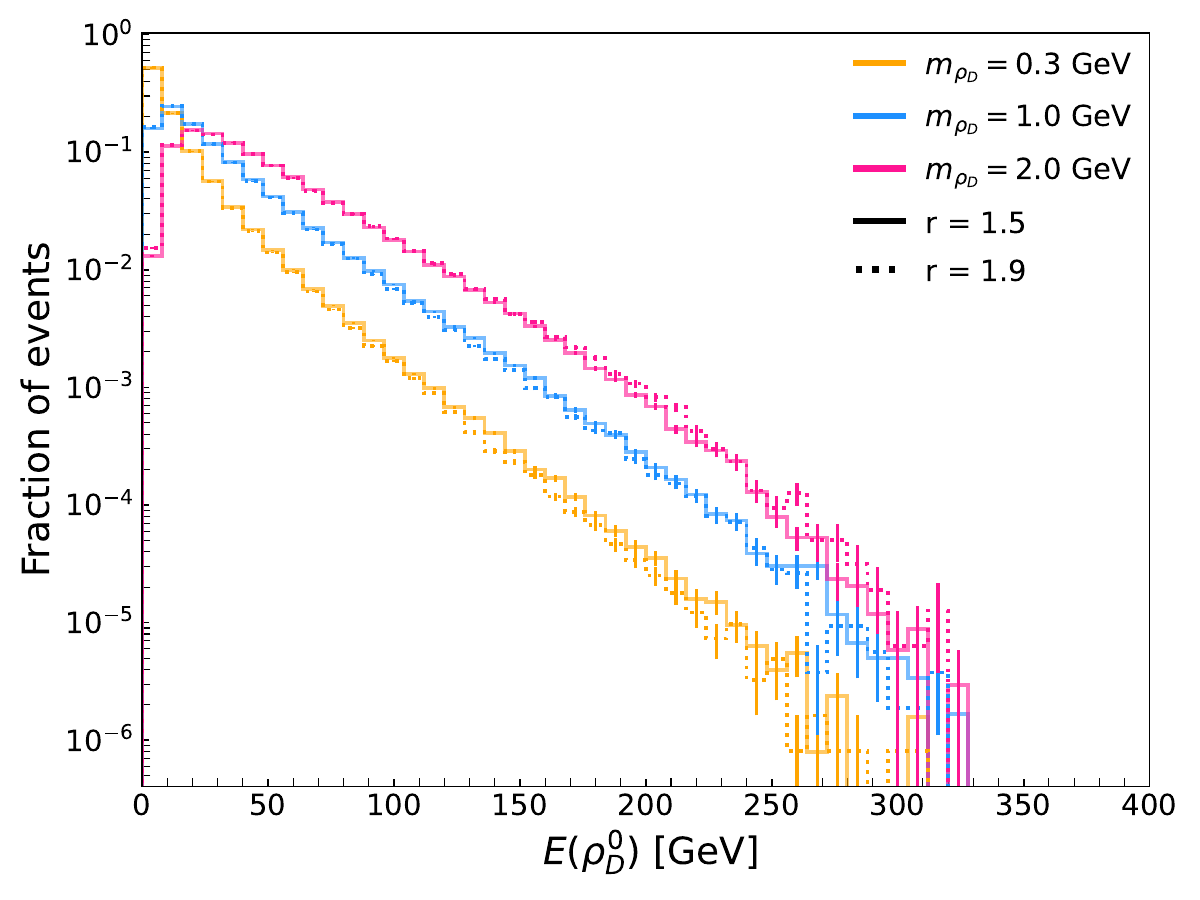}~
  \includegraphics[width=0.5\textwidth]{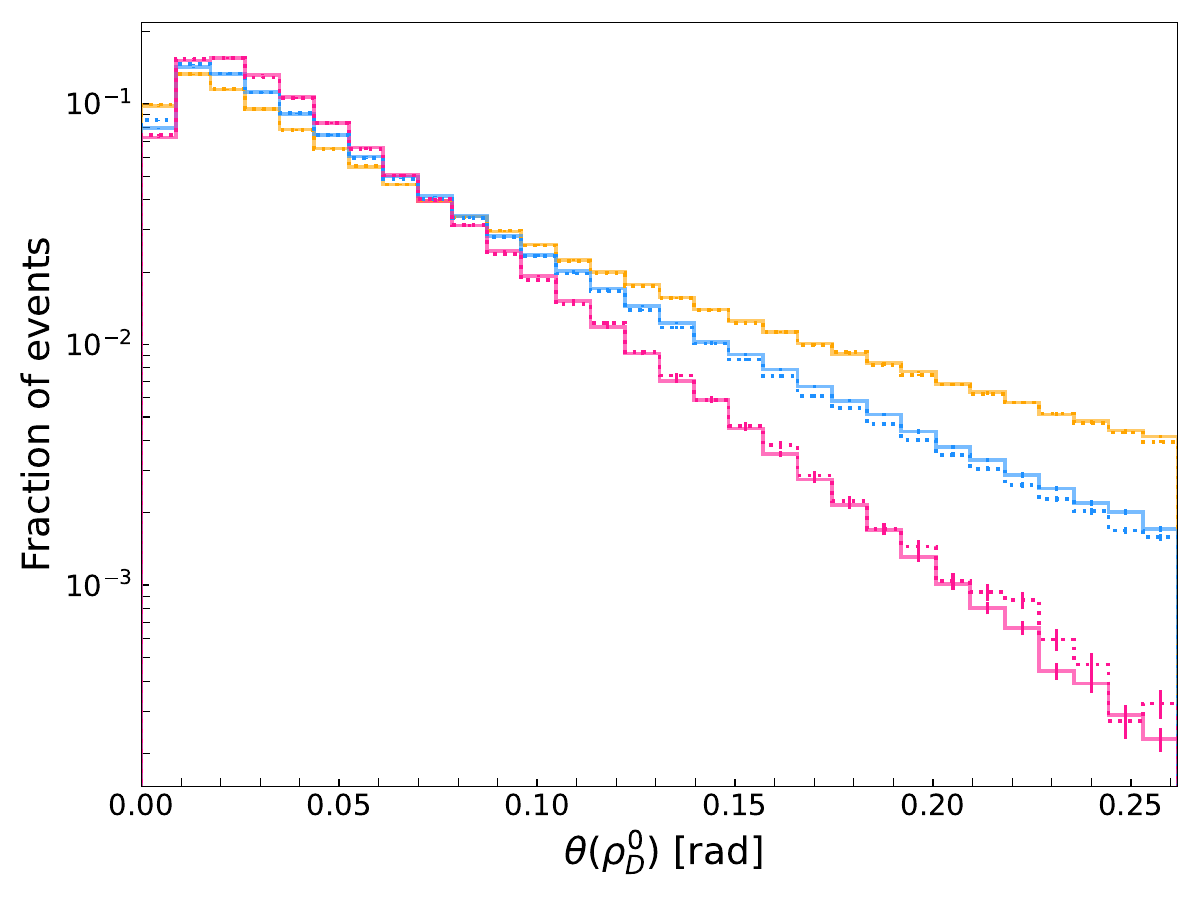}
  \vspace{-5mm}
\caption{%
 {Kinematic distributions of the dark rho mesons predicted for the SPS beam dump facilities. The panels show the energy (left) and polar angle (right) distributions of the produced neutral dark rho mesons. The solid lines show the distributions for $r=1.5$ and the dotted lines for $r=1.9$. The distributions are normalised to unity. Error bars indicate the standard Poisson uncertainties of the Monte Carlo samples.}}
 \label{fig:kinematic_distributions}
\end{figure*}

A dark shower results from the perturbative production of a pair of dark quarks via the effective interaction given in eq.~\eqref{eq:interactions}. For proton-beam dump experiments, the partonic centre-of-mass energy varies from event to event, but it can be significantly larger than the dark meson masses, such that a high multiplicity of dark rho mesons can be produced in the subsequent fragmentation and hadronisation processes. To determine these multiplicities as well as the distribution of energies and angles, we generate the hard process with \textsc{MadGraph v3.5.8} \cite{Alwall:2014hca} using a UFO model file created with \textsc{feynrules} and simulate the dark parton shower using the Hidden Valley module of \textsc{Pythia v8.313} \cite{Bierlich:2022pfr}. \footnote{We note here the existence of a \textsc{Herwig} hidden valley module~\cite{Kulkarni:2024okx}. It will be very interesting to simulate sub-GeV pion dark matter once the module becomes publicly available.}

In addition to the dark sector parameters discussed in section~\ref{sec:set-up}, \textsc{Pythia} requires a number of additional input parameters describing the evolution of the dark shower. The most important of these parameters is the relative probability of producing vector mesons, called \texttt{probVec}. A naive estimate based on spin degrees of freedom would suggest that the production of vector mesons should be three times as likely as for scalar mesons, corresponding to \texttt{probVec = 0.75}. However, since the dark rho mesons are slightly heavier, one expects the probability to be suppressed slightly. A recent study proposed a parametrisation for the resulting suppression as a function of the dark meson mass ratio $r$~\cite{Liu:2025bbc}. Based on these results, we take \texttt{probVec = 0.71} for $r = 1.5$ and \texttt{probVec = 0.68} for $r = 1.9$. All parameters relevant for our simulations are given in appendix~\ref{app:DetailsOnTheSimulatedSample}.

\begin{figure*}[t]
    \centering
    \includegraphics[width=0.49 \linewidth,clip,trim=0 0 50 0]
    {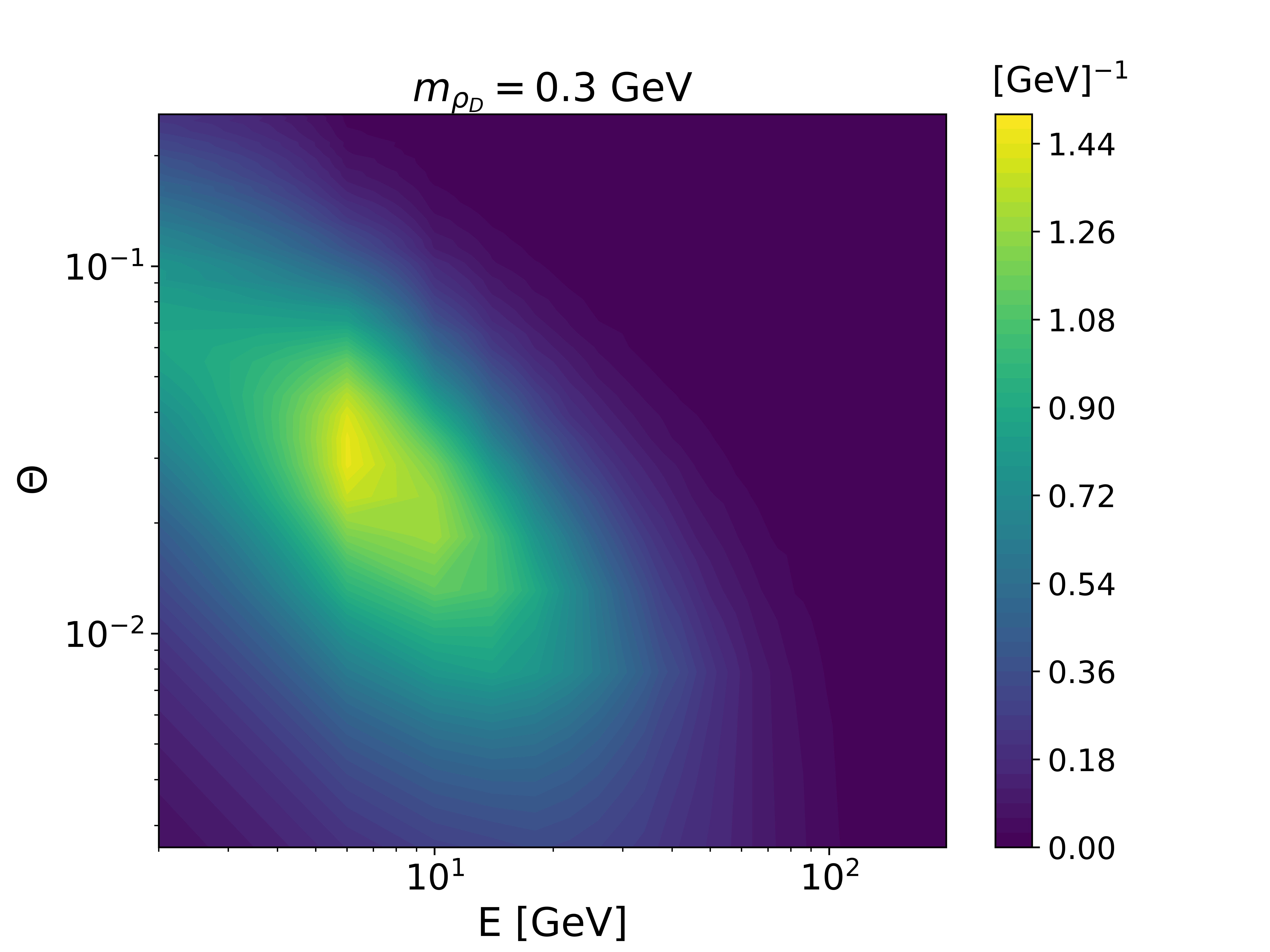}
    \hfill
    \includegraphics[width=0.49 \linewidth,clip,trim=0 0 50 0]{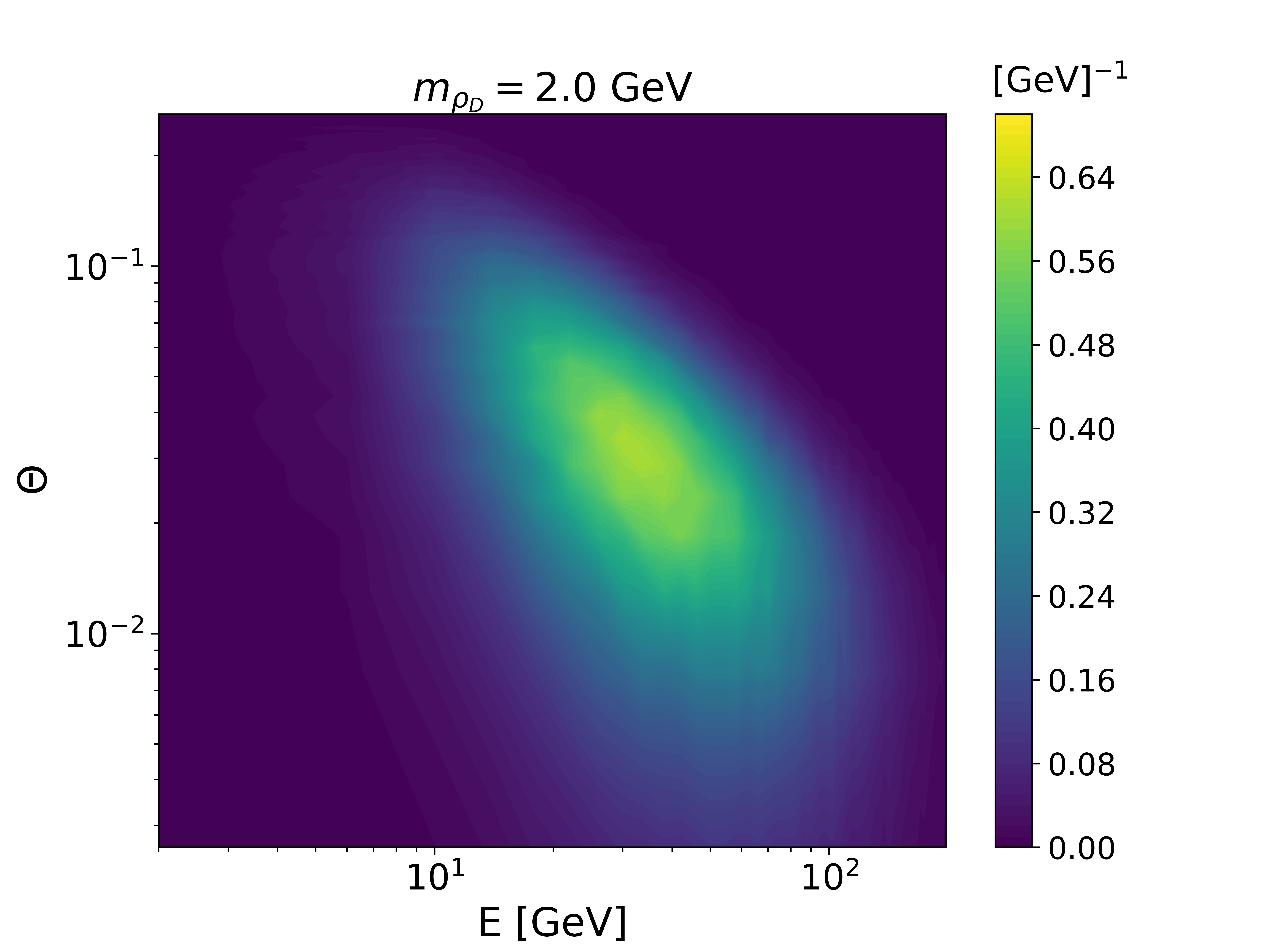}
    \vspace{-5mm}
    \caption{Normalised two-dimensional distribution of the energy and polar angle of the dark rho mesons predicted for the SPS beam-dump facilities for two different values of $\mrho$ and $r = 1.5$.}
    \label{fig:2d-densityplots}
\end{figure*}

Another subtlety concerns the simulation of sub-GeV dark mesons. Version 8.313 of \textsc{Pythia} is not intended to be used for such small masses and does not give reliable results.\footnote{While this work was being completed, a new version of \textsc{Pythia} capable of simulating sub-GeV dark showers was made public~\cite{py8history}.} However, there is a simple fix, which is to exploit the fact that the dark sector is fully characterised by a single dimensionful scale (such as $\mrho$) and a single dimensionless ratio (such as the dark meson mass ratio $r$). In other words, if $r$ is kept constant and $\mrho$ is rescaled by a factor $\chi$, the same rescaling factor also applies to $m_{\qd}$, $\mpi$, $\fpi$ and $\Ld$. If we simultaneously apply the same rescaling factor also to the centre-of-mass energy and to the suppression scale $\Lambda_\text{eff}$, we expect that dimensionless quantities such as dark meson multiplicities and angular distributions should remain invariant,\footnote{There is a small subtlety regarding the definition of the renormalisation and factorisation scale, see appendix~\ref{app:DetailsOnTheSimulatedSample} for details.} while all energies and momenta are rescaled by a factor of $\chi$ and the total cross section is rescaled by a factor of $\chi^{-2}$. Here, we assume that hadronisation parameters introduce no additional scale dependence. We have confirmed that within the \textsc{Pythia} Hidden Valley module, this is indeed the case for dark meson masses above 1 GeV and rescaling factors $1 < \chi < 10$. We can therefore also simulate dark meson masses below 1 GeV by first applying a sufficiently large rescaling factor, running the usual simulations, and finally applying the inverse rescaling to the total cross section and the energy distribution. In practice, we choose $\chi = 1 \, \mathrm{GeV} / \mpi$.

We show a few examples of the multiplicities and kinematic distributions in Figs.\ ~\ref{fig:multiplicities} and ~\ref{fig:kinematic_distributions}. The distribution of energy and polar angle refers to the lab frame, in which the incident proton has an energy of $400 \, \mathrm{GeV}$. As expected, we see larger multiplicities for smaller dark meson masses, which in turn means that each dark meson receives, on average, a smaller fraction of the available collision energy. To calculate the probability for a dark rho meson to decay in the SHiP detector, we furthermore need the two-dimensional distributions in energy and polar angle, which are shown in Fig.~\ref{fig:2d-densityplots} for two different cases. We see the expected anti-correlation between the two quantities. Particles that receive a large fraction of the energy of the incident beam tend to travel along the beam axis, while particles with lower energy are distributed over a wider angular range.

\begin{figure*}[t!]
    \centering
    \includegraphics[width=0.48\linewidth]{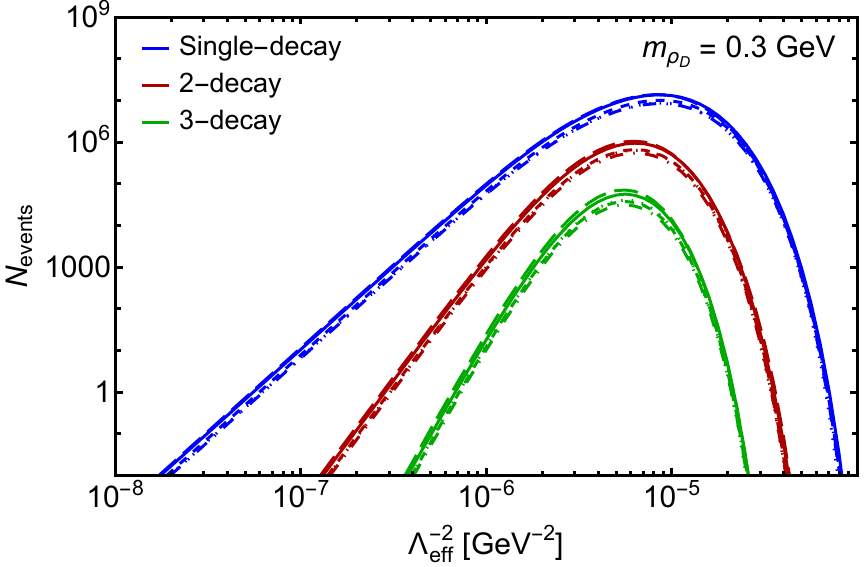}\hfill~\includegraphics[width=0.48\linewidth]{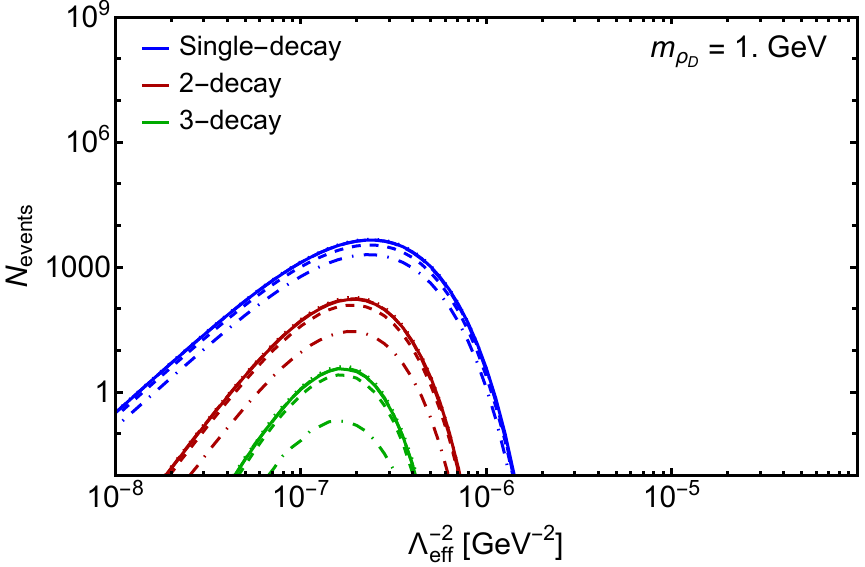}
    \caption{Impact of varying the dark-shower parameters on the event yields with one (blue), two (red), and three (green) decaying dark rho mesons per event at SHiP (15-year running time is assumed; see Sec.~\ref{sec:results} for details). Two benchmark masses are shown: $m_{\rho_D}=0.3~\mathrm{GeV}$ (left) and $1~\mathrm{GeV}$ (right) with the mass ratio fixed to $r = 1.5$ in both cases. Within each colour, the five line styles denote different parameter choices: \emph{solid} (baseline set-up with $\Ld = \Lambda_D^{\text{default}}$ as described in the main text and App.~\ref{app:DetailsOnTheSimulatedSample}); \emph{short-dash} $(\Lambda_D=\tfrac{1}{3}\Lambda_D^{\text{default}})$; \emph{long-dash} $(\Lambda_D=3\,\Lambda_D^{\text{default}})$; \emph{dash-dot} (\texttt{probVec}$=0.33$); \emph{dotted} (\texttt{probVec}$=0.75$).}
    \label{fig:uncertainties-impact-ship}
\end{figure*}

To better understand the impact of the uncertainties resulting from variations in the simulation set-up, it is not enough to show their impact on the production rate. The reason is that the number of events in a given experiment depends not only on the overall production yield but also on the kinematics of the dark rho mesons and their multiplicities per event, which all depend on the simulation set-up. Fig.~\ref{fig:uncertainties-impact-ship} shows the number of events at SHiP as a function of $\Lambda_{\text{eff}}$ for various choices of \texttt{probVec} and the scale $\Lambda_D$. We consider events with 1, 2, or 3 reconstructed decaying dark rho mesons per event, see Sec.~\ref{sec:analysis}. While for small values of $\mrho$ and the single-decay signature, the uncertainties have no substantial impact on our results, for larger dark rho meson masses and 2-decay or 3-decay signatures, the variation can be more than an order of magnitude.

\subsection{Dark-photon-like production}

\begin{figure*}
    \centering
    \includegraphics[width=\textwidth,clip,trim=0 5 0 2]{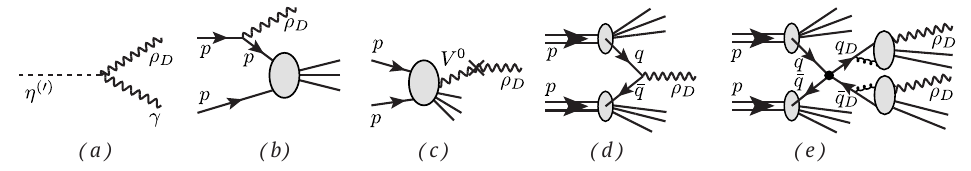}
    \caption{Illustration of the different production modes for dark rho mesons. Panels (a)--(d) show the various dark-photon-like production processes, namely (from left to right) decays of neutral mesons $\eta^{(\prime)}\to \rhod+\gamma$, proton bremsstrahlung, production via mixing with neutral mesons $V^{0}$, and Drell-Yan production. Panel (e) shows the production via dark showers specific to strongly interacting dark sectors.}\vspace{-2mm}
    \label{fig:production-diagrams-vector}
\end{figure*}

So far, we have focused on dark rho meson production in events with a total centre-of-mass energy much larger than the dark rho meson mass, such that the composite nature of the dark rho meson becomes relevant. In processes with lower energy, on the other hand, the
dark rho meson, when produced on-shell, behaves exactly like a fundamental particle with the coupling structure of a dark photon. This behaviour implies a number of additional production channels in hadronic collisions as shown in Fig.~\ref{fig:production-diagrams-vector}. These channels include: electromagnetic decays of SM mesons, proton bremsstrahlung; production in the fragmentation chain, when, effectively, the dark rho is produced similar to the vector mesons $V = \rho^{0},\omega,\phi$ in quark string fragmentation because of the $\rhod$--$V$ mixing; and the Drell-Yan process, with the hard process $q \bar{q} \to \rhod$. We neglect here the secondary production of $\rhod$ in electromagnetic cascades inside the thick targets of beam dump experiments, because the resulting dark rho mesons have tiny energies, typically well below the energy cuts on the decay products, which is $E_{\text{cut}} = 1\text{ GeV}$ for each decay product~\cite{Zhou:2024aeu,Kyselov:2024dmi}. 

The bremsstrahlung and fragmentation channels are significantly affected by the mixing of $\rhod$ with SM vector mesons $\rho^{0},\omega,\phi$ and their excitations, which may be understood in terms of vector meson dominance~\cite{Sakurai:1960ju,Fujiwara:1984mp,Ilten:2018crw}. This creates an overlap between their contribution to the $\rhod$ flux, leading to double counting. However, within the description used to calculate the probability of the bremsstrahlung -- the quasi-real approximation -- the particles are produced solely from initial state radiation~\cite{Altarelli:1977zs,Boiarska:2019jym} (see also Ref.~\cite{Foroughi-Abari:2024xlj}). The production in fragmentation, in contrast, includes both contributions. To remove this duplication, Ref.~\cite{Kyselov:2024dmi} only included production in final-state radiation when studying the fragmentation process.

The mixing with mesons induces sizeable theoretical uncertainties, in particular in the GeV mass range~\cite{Alimena:2025kjv}. This uncertainty is caused by the lack of experimental data on the properties of heavy excitations and their production cross-sections in inelastic proton collisions. In the case of the proton bremsstrahlung, an additional source of uncertainty is the quasi-real approximation, which introduces a proton with non-zero virtuality $q^{2}$. The virtuality distribution is however unknown and may heavily influence the production probability~\cite{Foroughi-Abari:2024xlj,Kyselov:2024dmi}. Altogether, the uncertainties in the bremsstrahlung flux may significantly exceed two orders of magnitude in the mass range $\mrho>1\text{ GeV}$~\cite{Foroughi-Abari:2024xlj,Kyselov:2024dmi,Kyselov:2025uez}. 

Given the resulting poor understanding of the bremsstrahlung production, in the calculations of the number of events, we do not include this production mode. Instead, in the production via fragmentation, we modify the description of Ref.~\cite{Kyselov:2025uez} by adding the production via initial state radiation. The resulting description does not account for the production via the mixing with heavy excitations of $\rho^{0},\omega,\phi$ states, and hence underestimates the production yield in the mass range $1\text{ GeV}<\mrho<2\text{ GeV}$.

\begin{figure}[t!]
    \centering
    \includegraphics[width=\linewidth]{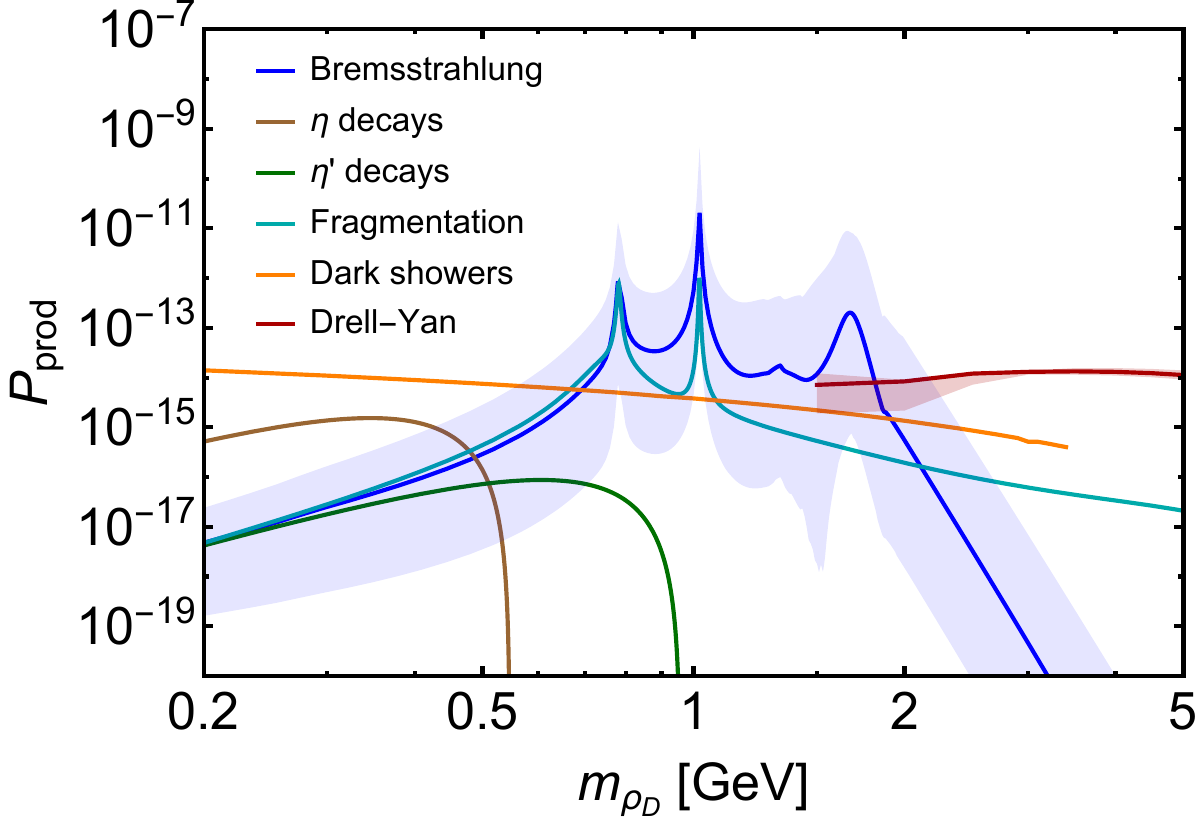}
    \caption{Production probabilities of dark rho mesons in the collision of the SPS proton beam with the Tungsten target corresponding to the SHiP experiment. The shaded regions show the theoretical uncertainties in the proton bremsstrahlung and Drell-Yan production mechanisms, see Refs.~\cite{Kyselov:2024dmi,Kyselov:2025uez} for details. Because of large theoretical uncertainties, we do not include the bremsstrahlung contribution. However, to avoid making the yield too conservative, in the fragmentation channel, we include the bremsstrahlung-mimicking production in the initial state radiation. For this plot we assume $r = 1.5$ and $\Lambda^{-2}_\text{eff} = 10^{-6} \, \mathrm{GeV}^{-2}$.}
    \label{fig:production-probabilities}
\end{figure}

We show the probability of producing a dark rho meson at SPS energies for the channels listed above in Fig.~\ref{fig:production-probabilities}. We find that production in dark showers dominates for $\mrho < m_\rho \approx 0.78 \, \mathrm{GeV}$. This is because the effective coupling $\epsilon$, which enters in the other production channels, is proportional to $\mrho^2$. On the other hand, the Drell-Yan production dominates for $\mrho > 2 \, \mathrm{GeV}$. In the intermediate regime, there is a competition between several different modes, including the production in dark showers.

In this work we only consider dark showers initiated by dark quark production from an SM quark initial state. In principle, SM meson decay and proton bremsstrahlung may also give rise to a dark shower if the momentum flowing into the dark sector is sufficiently large compared to $\Lambda_D$ (see Refs.~\cite{Contino:2020tix, Costa:2022pxv, Borrello:2025hal} for studies of inclusive production of dark sector particles via these channels). However, for the dark meson mass range considered in this work, only dark shower production from partonic SM initial states is relevant.

\section{Analysis methodology}
\label{sec:analysis}

In this section, we present the experiments that we consider and discuss how we calculate constraints and sensitivity projections. 

\subsection{SPS-based experiments}
\label{sec:sps}

We consider the following experiments located at the 400-GeV proton beam of the SPS at CERN: the past experiments BEBC~\cite{BEBCWA66:1986err,WA66:1985mfx} and CHARM~\cite{CHARM:1983ayi,CHARM:1985anb}; the currently running experiment  NA62~\cite{NA62:2017rwk,NA62:2023qyn,NA62:2025yzs}; and the approved beam-dump facility SHiP~\cite{SHiP:2015vad,SHiP:2025ows}, assuming the full 15-year running time with $N_{\text{PoT}} = 6\times 10^{20}$ protons on target. For NA62, we consider the beam-dump mode with two different values for the number of protons on target: $N_{\text{PoT}} = 1.4\times 10^{17}$, corresponding to the current constraints~\cite{NA62:2025yzs}, and $N_{\text{PoT}} = 1\times 10^{18}$, which is the full luminosity to be collected by 2026 and defines the ultimate sensitivity that can be achieved~\cite{Jerhot:2936260}. Details of the set-ups and selection criteria for the decay events are summarised in Ref.~\cite{Kyselov:2024dmi}. 

Possible signatures of the model include ``traditional'' displaced decays of one $\rhod$ per event, as well as ``$n$-decay'' events, in which one observes $n>1$ dark rho mesons decaying in the same event. To first approximation, the event rate of $n$-decays is suppressed compared to the single decay by a factor $P_{\text{decay}}^{n-1}$, where $P_{\text{decay}} \ll 1$ is the probability that a given dark rho meson decays inside the decay volume (see appendix~\ref{app:event-calc}). Given that the experiments that we consider operate(d) in a background-free environment, the single-decay events give the strongest exclusion limits and the best sensitivities, while the $n$-decay signature is subdominant. 

As was recently pointed out in Ref.~\cite{DallaValleGarcia:2025aeq}, however, the $n$-decay signature may play a crucial role in case a signal is observed at SHiP. The reason is that beam-dump experiments do not see the production vertex -- it is buried inside the target. Hence, their ability to distinguish between models relies mainly on the reconstruction of decay vertices. However, the phenomenology of $\rhod$ decays is the same as that of dark photons $V$. Hence, observing only single-decay events would not allow for model discrimination. The $n$-decay signature, on the other hand, could potentially confirm that the observed events originate from a strongly interacting dark sector. 

\begin{figure*}[t!]
    \centering
    \includegraphics[width=0.5\linewidth]{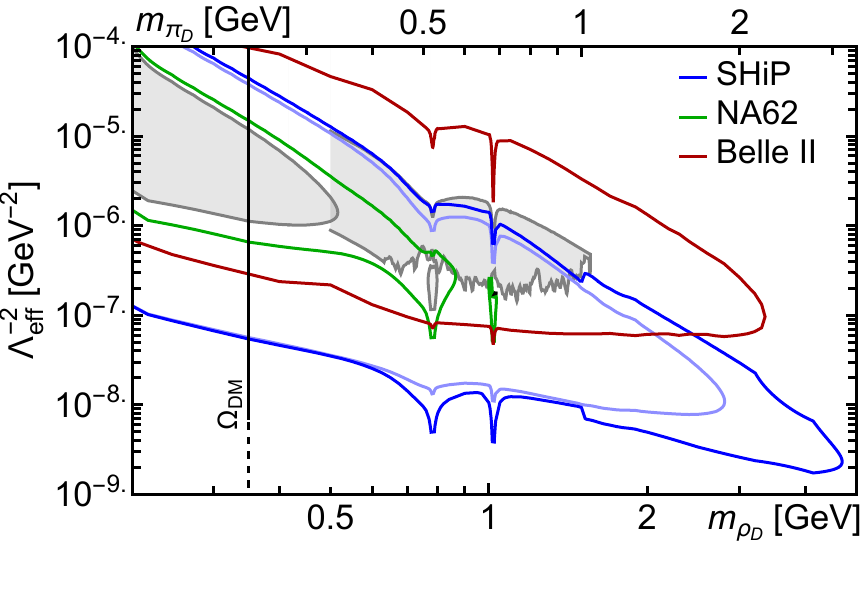}~\includegraphics[width=0.5\linewidth]{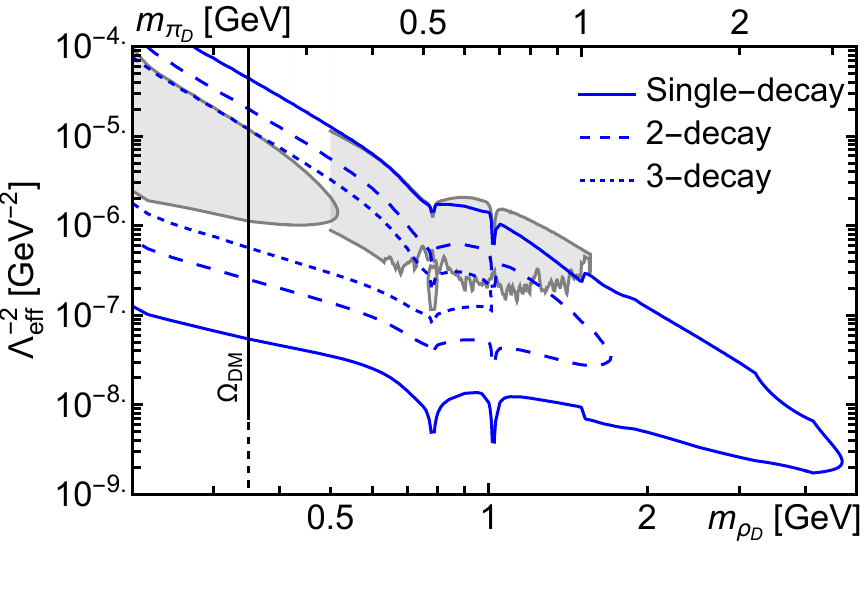}
   \caption{Experimental constraints and sensitivities in the parameter space of a strongly interacting dark sector. Left panel: exclusion limits and sensitivity projections coming from the signature of one decaying $\rhod$ per event. The gray region shows constraints coming from BEBC, CHARM, NA62 in the dump mode (assuming $N_{\text{PoT}} = 1.4\cdot 10^{17}$), and BaBar, while the coloured lines denote the sensitivities of SHiP, NA62 (with $N_{\text{PoT}} = 1\times 10^{18}$, to be collected by 2026), and Belle II (for $\mathcal{L} = 500 \, \mathrm{fb}^{-1})$. For SHiP, we show the contribution from the dark showers alone (the light blue line) and the full sensitivity (the solid line) for the expected 15 years of running time. The vertical black line indicates the dark pion mass for which the freeze-out of $3\pid \to \pid \rhod$ conversions results in the observed dark matter relic abundance~\cite{Bernreuther:2023kcg} (see main text for details). The relic density calculation assumes that the dark rho mesons are in equilibrium with the SM bath, which may not be the case for very small values of $\Lambda_\text{eff}^{-2}$, as indicated by the dotted line style. Right panel: the sensitivity of SHiP to events with one, two, and three decaying dark rho mesons observed per event.}
    \label{fig:beam-dump-detailed}
\end{figure*}

To calculate the event rate, we employ two different techniques:
\begin{enumerate}
\item A semi-analytic method as implemented in \textsc{Sens-Calc}~\cite{Ovchynnikov:2023cry}. This method takes as input tabulated angle-energy distribution $\mathrm{d}^{2}f/\mathrm{d}\theta \mathrm{d}E$, treating events with $n$ dark rho mesons as $n$ independent events. Accordingly, the total production cross-section for dark rho mesons is taken to be 
\begin{equation}
\sigma_{\text{prod}} = \sigma\times\langle N_{\rhod}\rangle,
\label{eq:production-cross-section}
\end{equation}
with $\sigma$ being the dark shower production cross-section and $\langle N_{\rhod}\rangle$ being the $\rhod$ multiplicity per event.
The semi-analytic approach only works for single-decay events and implicitly assumes that the contribution from di-decay events is negligible at the boundary of the sensitivity domain. We will see below (Fig.~\ref{fig:beam-dump-detailed}) that this assumption is well-justified, and hence the semi-analytical method is sufficient for the case of single-decay events. 
\item A Monte-Carlo event generator using importance sampling based on \textsc{EventCalc}, which is an add-on of \textsc{SensCalc}~\cite{DallaValleGarcia:2025aeq}. It uses the same basic set-up as \textsc{SensCalc}, but works directly with the $\rhod$ samples, processing them event by event. Unlike \textsc{Sens-Calc}, it is therefore able to compute also the rate of $n$-decay events. Further details on the implementation are summarised in App.~\ref{app:event-calc}.
\end{enumerate}

Of course, \textsc{EventCalc} can also be used to calculate bounds and sensitivities for single-decay events, thus providing an independent cross-check of the results obtained with \textsc{SensCalc}. We have performed this test for the case of the SHiP sensitivity projection, finding excellent agreement between the two approaches.\footnote{The implementation in \textsc{SensCalc} with the dark rho meson model is public and available on Zenodo~\cite{SensCalc-Zenodo} or \faGithub.
The module of \textsc{EventCalc} that calculates $n$-decay event rates for dark rho mesons may be provided upon request.}

\subsection{$B$ factories}

Just like in proton-proton collisions, dark showers can also be produced in electron-positron collisions via the effective interaction in eq.~\eqref{eq:interactions}. $B$ factories, operating at a centre-of-mass energy of $\sqrt{s} \approx 10.6 \, \mathrm{GeV}$, have an excellent track reconstruction and can therefore be used to search for displaced vertices from dark rho meson decays. Ref.~\cite{Bernreuther:2022jlj} performed both a reinterpretation of constraints from BaBar~\cite{BaBar:2015jvu} and a sensitivity projection for Belle II based on the search strategy proposed in Ref.~\cite{Duerr:2019dmv}. However, Ref.~\cite{Bernreuther:2022jlj} did not use lattice results to relate $\mrho$, $\mpi$ and $\fpi$ and simply set $g_{\rho \pi} = 1$. 
Moreover, the Hidden Valley module of \textsc{Pythia} was used to simulate dark showers for sub-GeV dark meson masses without the rescaling procedure described in Sec.~\ref{sec:phenomenology}.\footnote{$B$-factories prospects for strongly interacting dark matter scenarios were also discussed in~\cite{Hochberg:2017khi,Davighi:2024zip,Davighi:2025awm}.}

Here we provide an updated version of these results, using the exact same model and \textsc{Pythia} settings as for the proton beam-dump experiments. We implement the trigger requirements and event selection as described in Ref.~\cite{Bernreuther:2022jlj}. For Belle II, we consider dark rho meson decays to electrons if they happen either at a distance $0.2\,\mathrm{cm} \leq r \leq 0.9\,\mathrm{cm}$ or $17\,\mathrm{cm} \leq r \leq 60\,\mathrm{cm}$ from the beam axis in the radial direction, and to muons, pions or kaons over the full range $0.2\,\mathrm{cm} \leq r \leq 60\,\mathrm{cm}$. We emphasise that it is not clear whether this search can be background-free, such that we refrain from estimating the sensitivity for the total planned Belle II luminosity and instead show projections for $500\,\mathrm{fb}^{-1}$. 

\subsection{LHC}

The model that we consider also predicts the production of dark showers at the LHC. As for SPS-based experiments, the displaced vertices resulting from dark rho meson decays resemble those of dark photons and can therefore be explored with similar search strategies. The general expectation is that due to the larger boost factors and the more compact detectors, LHC experiments are sensitive to smaller dark rho meson lifetimes than what can be explored with proton beam-dump experiments. This expectation was confirmed explicitly in Ref.~\cite{Bernreuther:2022jlj}, which performed a reinterpretation of a search for displaced di-muon resonances at LHCb~\cite{LHCb:2020ysn}. However, as also pointed out in Ref.~\cite{Bernreuther:2022jlj}, the effective description of eq.~\eqref{eq:interactions} is not valid at LHC energies, because the $Z'$ boson can be produced on-shell. As a result, cross sections and distributions depend on $m_{Z'}$, $\kappa$, and $e_d$ in a more complicated way, making a direct comparison with the sensitivity of low-energy experiments more difficult.

In the present work, we restrict ourselves to the effective interaction between quarks and dark quarks and therefore do not consider possible constraints and sensitivities of LHC-based experiments, which are expected to cover different regions of parameter space compared to SPS-based experiments. We note, however, that there are various ideas to extend the sensitivity of LHC-based experiments to longer lifetimes, both using larger parts of the existing detectors (for example, the downstream trackers of LHCb~\cite{Gorkavenko:2023nbk} or the muon detectors of ATLAS~\cite{ATLAS:2018tup,ATLAS:2019jcm} and CMS~\cite{Liu:2025bbc,CMS:2021juv}) and constructing dedicated detectors at larger distances, see Ref.~\cite{PBC:2025sny}. Studying the sensitivity of these experiments to strongly interacting dark sectors at the sub-GeV scale is an exciting prospect for future work.

\section{Constraints and sensitivities}
\label{sec:results}

The constraints and sensitivities of various experiments are shown in the left panel of Fig.~\ref{fig:beam-dump-detailed} for $r = 1.5$. The corresponding results for $r = 1.9$ are found to be nearly identical when expressed in terms of $\mrho$ and $\Lambda_\text{eff}$, with a shift in the corresponding values of $\mpi$. The exclusion regions from past and current beam dump experiments are limited to the range $\mrho\lesssim 0.5\text{ GeV}$, mainly because of the limited beam intensity and geometric acceptance: the decay volumes cover a tiny amount of the solid angle in the far-forward direction. For heavier dark rho mesons, the model-agnostic long-lived particle search at BaBar excludes a part of parameter space reaching to masses of up to $\mrho \approx 1.6$~GeV. However, its sensitivity is limited by the trigger criteria, which require at least three tracks with $p_T > 0.12$~GeV, corresponding to at least two observed $\rhod$ decays in the same event (or one observed $\rhod$ decay with additional tracks from ISR). Note that efficiency tables for this search are only available for LLPs above 500~MeV, preventing a reinterpretation for smaller masses. For NA62, the event rate is significantly suppressed by the acceptance of decay products. With the full statistics to be collected at NA62, the sensitivity may be extended to $\mrho \sim 0.7\text{ GeV}$.

Thanks to its huge beam intensity and large geometric acceptance, SHiP will be able to probe dark rho meson masses up to $\mrho \sim 5\text{ GeV}$. As expected, the sensitivity reach is determined by the single-decay signature. Dark rho meson production via dark showers significantly contributes to the mass range $\mrho\lesssim 3\text{ GeV}$, while at larger masses the sensitivity is dominated by Drell-Yan production (see Fig.~\ref{fig:production-probabilities}). The probed parameter region is highly complementary to the one explored by Belle II, which is sensitive to shorter decay lengths and hence larger effective couplings. 

Both SHiP and Belle~II probe the sub-GeV range of dark meson masses that are required for the freeze-out of $3 \to 2$ annihilations to yield the correct relic density for dark pion dark matter. In particular, dark matter freeze-out via $3\pid \to \pid \rhod$ conversions predicts a dark pion mass of $\mpi \approx 0.33\,\mathrm{GeV}/\nf^{2/3}$, nearly independent of other dark sector parameters~\cite{Bernreuther:2023kcg}. For $\nf=2$, this relic density target is given by $\mpi\approx 0.21$~GeV (corresponding to $\mrho \approx 0.32$~GeV for $r=1.5$) and is indicated by the vertical black line in Fig.~\ref{fig:beam-dump-detailed}.

The relic density prediction does not depend on $\Lambda_\text{eff}$ as long as the (inverse) decays of the dark rho meson are fast enough to keep the dark rho mesons in equilibrium with the SM bath during dark matter freeze-out. This is the case if $\Gamma_{\rhod}/H(T_f) \gtrsim \mpi/T_f$, where $H$ denotes the Hubble rate and $T_f \approx \mpi/20$ the temperature at which dark matter decouples. For $m_\rhod = 0.32$~GeV, this equilibrium condition requires that $\Lambda_\text{eff}^{-2} \gtrsim 7 \times 10^{-9}$~GeV$^{-2}$. This represents a lower bound on the parameter space in which $3\pid \to \pid \rhod$ annihilations can set the dark matter relic density without being limited by the interaction strength between the dark sector and the SM. For this reason, the relic density line is dotted for smaller values of $\Lambda_\text{eff}^{-2}$  in Fig.~\ref{fig:beam-dump-detailed}.

In addition to the single-decay signature used to derive the sensitivity in the left panel of Fig.~\ref{fig:beam-dump-detailed}, SHiP also has the potential to explore the signatures with two and three decaying dark rho mesons per event. As shown in the right panel of Fig.~\ref{fig:beam-dump-detailed}, SHiP can expect to see 2-decay events and 3-decay events for dark rho meson masses up to  $\mrho \lesssim 2\text{ GeV}$ and $\mrho \lesssim 1\text{ GeV}$, respectively. A large fraction of this parameter space is compatible with all current constraints, although most of it will be probed in the coming years by NA62 and Belle II. In other words, SHiP would be able to follow up on a signal in one of these experiments to provide more details on the underlying structure of the dark sector.

Observing two decaying dark rho mesons per event is already enough to rule out the minimal dark photon model, in which the probability to produce multiple dark photons in the same event is vanishingly small. However, there exist non-minimal dark photon models, which allow for the pair-production of dark photons (see, for example, Ref.~\cite{Curtin:2014cca}). In this case, there are two ways to discriminate between the models. The first way is to search for events with three or more decaying dark rho mesons, which are of course even less likely than 2-decay events.\footnote{We note that in dark photon models featuring a dark Higgs mechanism~\cite{Schabinger:2005ei,Batell:2009yf} it is also possible to produce three dark photons in a single event through the dark Higgs-strahlung process $V^\ast \to V h_d^{(*)}$ followed by $h_d^{(\ast)} \to V V$. The rate for this process is, however, expected to be substantially suppressed compared to single dark photon production~\cite{Duerr:2017uap}, enabling discrimination between the two models based on the relative rates of single-decay and multi-decay signatures.}

\begin{figure}[t!]
    \centering
    \includegraphics[width=\linewidth]{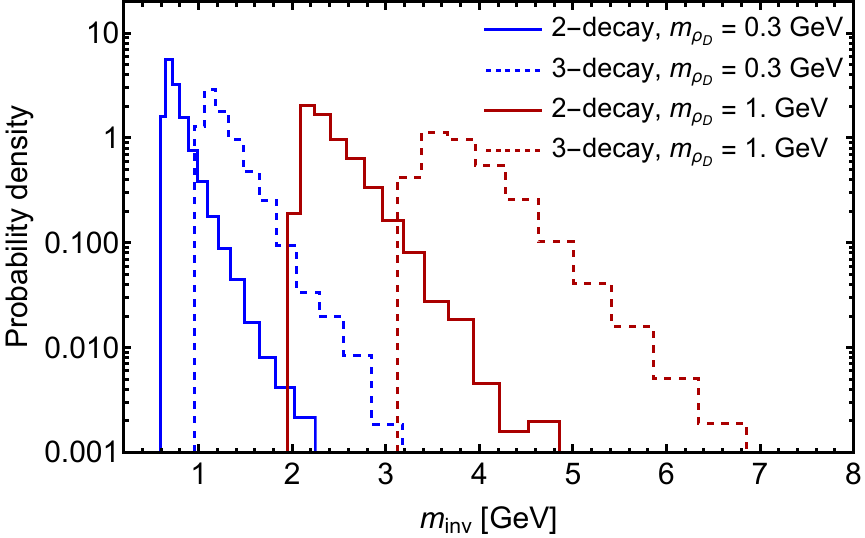}
    \caption{Total invariant mass of the dark rho mesons produced in dark showers and decaying inside the decay volume of SHiP, with all the decay products passing the acceptance criteria (see main text for details). We show the cases of two (solid) and three (dashed) decaying dark rho mesons for two masses: $\mrho=0.3$ GeV (blue) and $1$ GeV (red).}
    \label{fig:minv}
\end{figure}

A promising alternative is to measure the total invariant mass $m_{\text{inv}}$ in the events with $>1$ decaying dark rho mesons (see also Refs.~\cite{Han:2007ae,DallaValleGarcia:2025aeq}). Examples for such distributions are shown in Fig.~\ref{fig:minv}. We find that the invariant mass distribution is rather broad and extends from the threshold $m_{\text{thr}} = n\cdot \mrho$ up to larger values. This result immediately enables us to differentiate the dark rho meson model from more minimal extensions of the dark photon model that feature pair production through the decays of a heavier state. For example, if a dark photon pair is produced in the decay $B_{s} \to 2V$, the reconstructed invariant mass peaks at the mass of the $B_s$ meson. Broader invariant mass distributions could result from the decay $B\to X_{s/d}+2V$, but in this case, the invariant mass of the dark-photon pair is bounded from above by the $B$ meson mass, while no such limitation exists for the case of dark rho mesons.
We note that similar measurements may also be possible at Belle II, if one observes events with two displaced vertices that may originate from decaying dark rho mesons.

\section{Conclusions}
\label{sec:conclusion}

Dark sectors with new strong interactions provide attractive dark matter candidates in the form of dark mesons and predict novel dark shower signals at accelerator experiments. While the sub-GeV range of dark meson masses motivated by dark matter freeze-out via $3 \to 2$ annihilations is difficult to probe at high-energy colliders, beam dump experiments are a natural place to search for light particles with long lifetimes. In this work, we have studied the sensitivity of beam dump and $e^+e^-$ collider experiments to strongly interacting dark sectors at the GeV and sub-GeV scale, with a focus on the recently approved SHiP facility and other experiments at the CERN SPS.

For concreteness, we have considered a QCD-like dark sector with two flavours of dark quarks in the fundamental representation of a new $SU(3)$ gauge group. After dark sector confinement, these dark quarks form stable pseudo-Nambu-Goldstone bosons (dark pions $\pid$) and unstable vector mesons (dark rho mesons $\rhod$). The dark sector is coupled to SM particles via an effective interaction originating from a heavy (integrated out) $Z'$ boson. The dark sector is then fully described by three parameters: the masses $\mpi$ and $\mrho$ and the suppression scale of the effective operator $\Lambda_\text{eff}$. We set all other parameters of the strongly interacting dark sector consistently based on lattice results.

In the case that $\mrho < 2 \mpi$, which is motivated by cosmology and astrophysics, the dark rho mesons can decay into SM particles and thus lead to visible experimental signatures. At the same time, the effective interaction mediating their decay also allows for the production of dark rho mesons in $p$-$p$ collisions, either via the hadronisation of dark quarks in dark showers or via standard dark-photon-like production modes. We take all production mechanisms into account simultaneously and find that dark showers dominate dark rho meson production at the SPS at CERN for $\mrho \lesssim 0.8$~GeV and significantly contribute up to $\mrho \approx 3$~GeV (see Fig.~\ref{fig:production-probabilities}).

Based on simulated events with long-lived dark rho mesons, we have determined constraints from past beam dump experiments and the prospective sensitivity of SHiP using \textsc{SensCalc}. We find that SHiP can probe a large currently unconstrained parameter space reaching up to $\mrho \approx 5$~GeV and $\Lambda_\text{eff} \approx 20$~TeV, and is highly complementary to Belle~II, for which we have calculated updated dark shower sensitivity projections (see Fig.~\ref{fig:beam-dump-detailed}).

Both SHiP and Belle~II are sensitive to the dark meson masses that are predicted if the relic abundance of dark matter is set by the freeze-out of $3 \to 2$ annihilations, in particular $3\pid \to \pid \rhod$ conversions. Moreover, we find that SHiP will probe nearly the entire range of $\Lambda_\text{eff}$ for which this mechanism of producing the relic abundance is not limited by the strength of the interaction between the dark sector and the Standard Model.

In addition, there is a sizeable parameter space in which SHiP is expected to see events with multiple observed dark rho meson decays in the same event. We find that the reach for events with two observed decays extends up to $\mrho \approx 2$~GeV and for events with three observed decays up to $\mrho \approx 1$~GeV. In case of a discovery, these events will be crucial to distinguish strongly interacting dark sectors from dark photon models, either by the sheer multiplicity of decays or the invariant mass spectrum of all decay products (see Fig.~\ref{fig:minv}).

Throughout this work, we have described the interaction between the dark sector and the SM via an effective operator. Its UV completion may be probed at the LHC. Although the sub-GeV masses of dark mesons motivated by cosmology and their associated large decay lengths are outside the sensitivity reach of typical long-lived particle searches at the LHC, this hurdle may be overcome with searches beyond the tracker and dedicated detectors further away from the interaction point.

\acknowledgments

We thank Torben Ferber, Chris Hearty and Joerg Jaeckel for helpful discussions. EB is supported, in part, by the US National Science Foundation under Grant PHY-2210177. NH and FK acknowledge funding from the Deutsche Forschungsgemeinschaft
(DFG) through Grant No. 396021762 -- TRR 257.  SK is supported by the FWF project number P 36947-N. SK thanks the Institute for Theoretical Particle Physics at the Karlsruhe Institute for Technology for hospitality and support during parts of this work.

\appendix

\section{Details on the dark shower simulations}
\label{app:DetailsOnTheSimulatedSample}

For the dark shower simulations, we consider the range $0.2 \, \mathrm{GeV} \leq \mrho \leq 3.4 \, \mathrm{GeV}$ and the two mass ratios $r = \mrho / \mpi = 1.5$ and $1.9$. The other dark sector parameters are set following Ref.~\cite{Albouy:2022cin} as
\begin{align}
\Ld & =\frac{5\mpi}{12}  \sqrt{r^2 - 1.5} \\
m_{q_D\text{,curr}} &=\frac{4}{121} \frac{\mpi^2}{\Ld} \\  
m_{q_D\text{,const}} & =m_{q_D\text{,curr}} + \Ld
\end{align}
From these results and the KSRF relation, it follows that $g_{\pi\rho} \approx 5.7$ for both values of $r$. 

For the dark shower simulation in \textsc{Pythia}, we follow Ref.~\cite{Liu:2025bbc} and set \texttt{probVec} to 0.71 for $r = 1.5$ and to 0.68 for $r = 1.9$. 
We set \texttt{separateFlav = on} and set \texttt{probKeepEta1 = 0} to suppress the production of heavier dark mesons. Finally, we set \texttt{setLambda = on} for \textsc{Pythia} to use the provided value of $\Ld$ to calculate the running of the dark sector gauge coupling. 

For the simulation of the hard process, we also need to specify the value of $\Lambda_\text{eff}$, even though the results can trivially be rescaled to a different value. Concretely, we simulate $m_{Z'} = 1 \, \mathrm{TeV}$, $\kappa = 10^{-3}$ and $e_D = 1$, corresponding to $\Lambda_\text{eff} = 57.5 \, \mathrm{TeV}$. We set the renormalisation scale to the partonic centre-of-mass energy.
We use the NNPDF23\_nlo\_as\_0119 pdf set \cite{Buckley:2014ana}, setting the factorisation scale to $10\,\mathrm{GeV}$. Using a fixed factorisation scale ensures that the proton pdfs do not change when we apply the rescaling procedure discussed in section~\ref{sec:phenomenology}. We have checked that varying the factorisation scale in the range 2--20 GeV changes the total cross section by less than 20\% and does not significantly affect kinematic distributions. Finally, we require a minimal partonic centre-of-mass energy $\sqrt{\hat{s}_\text{min}} = 2 \mpi$ to exclude the Drell-Yan production of a single dark rho meson, which we include explicitly in the dark-photon-like production modes.

We generate $10^6$ events for each mass point of $\mrho$ between 0.2 GeV and 3.4 GeV in increments of 100 MeV. Additionally, we include mass points at $\mrho=0.78 \text{ and }1.02$ GeV to better capture the kinematics of the $\rhod$ around the masses of the SM $\rho,\omega$ and $\phi$ mesons, which mix with the dark rho meson and resonantly enhance its decay modes. 

We also consider events containing not only one but also two or three decaying dark rho mesons. Because such topologies are rare, 1-to-1 sampling would suffer from large statistical fluctuations. To increase the effective statistics, after simulating the $\rhod$ production, we oversample the classes with exactly two and three produced dark rho mesons by duplicating events until the sample sizes satisfy
$N_{\text{prod},2\rhod} > 5\times 10^{4}$ and $N_{\text{prod},3\rhod} > 3\times 10^{4}$. 
To further reduce variance, we apply an additional random rotation about the beam axis by an angle $\phi \in (-\pi,\pi)$. This step is nontrivial, since axial symmetry is broken in beam-dump experiments by both the geometry and the spectrometer magnetic field.

At the stage where displaced decays of the dark rho mesons are generated, these oversampled events are re-used with per-copy weights chosen to keep the normalization unbiased. Specifically, if events in the class with $k\in\{2,3\}$ produced dark rho mesons are replicated $c_k$ times to reach the target $N_{\text{prod},\,k\rhod}$, each copy is assigned weight
$w_k = 1/c_k$ (and $w_1=1$ for the $k=1$ class).

\section{Monte-Carlo sampling with \textsc{EventCalc}}
\label{app:event-calc}

\begin{figure}[t!]
    \centering
    \includegraphics[width=\linewidth]{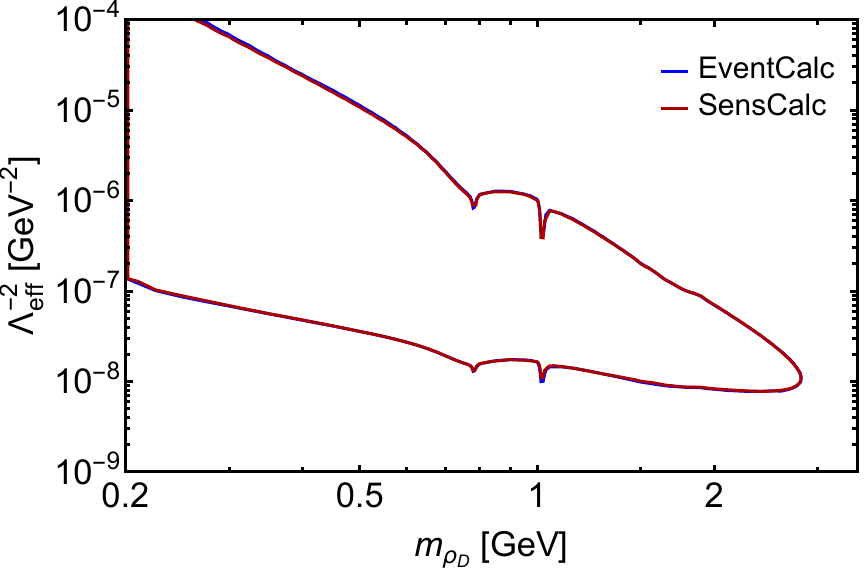}
    \caption{The comparison of the iso-contours with $N_{\text{events}}^{(1)}=2.3$ at the SHiP experiment as calculated using the Monte-Carlo sampler from Sec.~\ref{app:event-calc} (Eq.~\eqref{eq:nevents-monte-carlo}) and using the implementation of the dark rho meson model in \textsc{SensCalc}. Baseline selection criteria for decay events, summarised in Sec.~\ref{sec:analysis}, are used.}
    \label{fig:event-calc-vs-sens-calc}
\end{figure}

To compute the rate of events with $m$ decaying dark rho mesons in \textsc{EventCalc}, we have employed the following procedure. Consider an event $i$ with $n_{i}$ produced dark rho mesons:
\begin{enumerate}
    \item For each $\rhod$, we sample the decay vertices inside the SHiP decay volume, provided that the trajectory of the given $\rhod$ intersects the volume. The corresponding geometric weight, $\omega_{\text{dec.vol}}$, is $1$ in the case of intersection and 0 otherwise. The $z$ position is sampled using the inverse CDF from the differential decay probability 
    \begin{equation}
    \frac{dP_{\text{decay}}}{dz} = \frac{\exp\left[-\frac{z}{l_\rhod\cos(\theta)}\right]}{l_\rhod\cos(\theta)},
    \end{equation}
    where $l_\rhod=c\tau_{\rhod}\sqrt{\gamma^{2}-1}$ is the decay length of the decaying $\rhod$, with being its $\gamma$, and $\theta$ is the polar angle of the $\rhod$. Then, we calculate the decay probability:
    \begin{align}
        P_{\text{decay}} = & e^{-\tfrac{z_{\text{min}}}{l_{\rhod}\cos(\theta)}} -e^{-\tfrac{z_{\text{max}}}{l_{\rhod}\cos(\theta)}} 
    \end{align}
   with $z_{\text{min/max}}$ being the coordinates of the boundaries of the decay volume of the given experiment.
    \item For each of the decayed dark rho mesons, we sample the phase space of its decay products using the approach adopted in \textsc{SensCalc}: computing their 4-momenta at the dark rho meson's rest frame and then boosting to the lab frame. For decays into jets, we use the pre-tabulated phase space showered and hadronised in \textsc{PYTHIA8}~\cite{Bierlich:2022pfr}. Only the decay channels visible in the given experiment are considered; their total branching ratio is defined as $\text{Br}_{\text{vis}}$.
    \item Having simulated the decay products, we calculate their acceptance $\epsilon_{\text{dec}}$. The event is accepted ($\epsilon_{\text{dec}} = 1$) if the trajectories of the decay products intersect the detector and if the particles satisfy various cuts, such as energy cut, mutual spatial separation cut (in the electromagnetic calorimeter), transverse impact parameter cut (for pairs of particles), etc.
\end{enumerate}
The $k$th dark rho meson from the $i$th event has the following weight:
\begin{equation}
    \omega_{i,k} = \omega^{i,k}_{\text{dec.vol}}\times P_{\text{decay}}^{i,k}\times \text{Br}_{\text{vis}}\times \epsilon^{i,k}_{\text{dec}},
\end{equation}
Given these weights, it is possible to calculate the rate of the events with $m$ decaying dark rho mesons:
\begin{align}
    N_{\text{events}}^{(m)} = & N_{\text{PoT}}\cdot\frac{\sigma_{\geq 1 \rhod}}{\sigma_{pN}} \times \frac{1}{N_{\text{sample}}}\sum_{i = 1}^{N_{\text{sample}}}h(n_{i} - m) \nonumber \\
    &\times \sum_{\{m\}}\prod_{k\in \{m\}}\omega_{i,k}\times \prod_{t\notin \{m\}}(1-\omega_{i,t})
    \label{eq:nevents-monte-carlo}
\end{align}
Here, $\sigma_{pN} = 51\ A_{\text{target}}^{-0.29}\text{ mb}$ is the proton interaction cross-section per nucleon for the experiment's target, $\sigma_{\geq 1\rhod}$ is defined around Eq.~\eqref{eq:production-cross-section}, $N_{\text{sample}}$ is the size of the simulated sample, and $h(x)$ is the step function. The internal summation in Eq.~\eqref{eq:nevents-monte-carlo} is performed over distinct subsets $\{m\}$, each containing $m$ dark rho mesons.

As a cross-check of the framework, we have compared its predictions for SHiP with the calculation of the single-decays event rate within \textsc{SensCalc}. As discussed in the main text (Sec.~\ref{sec:analysis}), the latter uses $\sigma_{\geq 1\rhod}\times \langle N_{\rhod} \rangle$ as the production cross-section, and splits each event with $m>1$ of $\rhod$ produced mesons into $m$ independent events with 1 produced meson to obtain the angle-energy distribution of $\rhod$ mesons in these events.

Calculating the event rate involves not only sampling kinematics of the produced $\rhod$s, but also sampling their decay vertex position within the SHiP decay volume, simulating their decays, propagating the decay products through the detector, and calculating the decay products acceptance. Therefore, comparing the event rates, one would comprehensively test the routine. The perfect match within the Monte-Carlo noise is only possible if all the simulation steps are implemented correctly. 

Fig.~\ref{fig:event-calc-vs-sens-calc} shows the comparison of iso-contours $N_{\text{events}}=2.3$ in the parameter space $\mrho-\Lambda_{\text{eff}}^{2}$. The agreement in terms of $\Lambda_{\text{eff}}^{2}$ is within a few percents independently of the mass and lifetime of $\rhod$, which translates to the agreement within 10\% in terms of the number of events (as at the lower bound of the sensitivity $N_{\text{events}}\propto \Lambda_{\text{eff}}^{-8}$). This corresponds to the intrinsic error in the semi-analytic approach adopted in \textsc{SensCalc}, which comes from the discretization of the decay volume (see Ref.~\cite{Ovchynnikov:2023cry} for details).

\bibliography{bib.bib}

\end{document}